\documentclass[aps,prl,twocolumn,superscriptaddress,showpacs]{revtex4}
\usepackage{epsfig}
\newcommand{\ppbar} {p\overline{p}}
\newcommand{\pbinv} {{\mathrm{pb}^{-1}}}

\newcommand{\goto} {\rightarrow}
\newcommand{\epem}   {e^+e^-}
\newcommand{\mpmm}   {\mu^+\mu^-}
\newcommand{\tptm}   {\tau^+\tau^-}
\newcommand{\lplm}   {\ell^+\ell^-}

\newcommand{\MET}     {\not\!\!\!E_T}
\newcommand{\METx}    {\not\!\!\!E_{T,x}}
\newcommand{\METy}    {\not\!\!\!E_{T,y}}
\newcommand{\ET}      {E_T}
\newcommand{\pt}      {p_T}
\newcommand{\Ehad}    {E_{\mathrm{had}}}
\newcommand{\Eem}     {E_{\mathrm{em}}}

\newcommand{\Gammatot} {\Gamma^{\mathrm{tot}}_W}
\newcommand{\Vcs}      {|V_{cs}|}

\newcommand{\fermilab} {{Fermilab}}
\newcommand{\tevatron} {{Tevatron}}
\newcommand{\cdf} {{\sc CDF}}
\newcommand{\dzero} {{\sc D}\O}
\newcommand{\pythia}   {{\sc Pythia}}
\newcommand{\geant}    {{\sc Geant}}

\newcommand{\cteqsix}   {{ CTEQ6M}}
\newcommand{\mrstpdf}   {{ MRST}}

\newcommand{\stat}   {{\mathrm{(stat)}}}
\newcommand{\syst}   {{\mathrm{(syst)}}}
\newcommand{\lumerr} {{\mathrm{(lum)}}}

\newcommand{\etal} {{\em et al.}}

\begin{document}
\title{First Measurements of Inclusive $W$ and $Z$ Cross Sections\\ 
from Run~II of the Tevatron Collider}
%
%

\affiliation{Institute of Physics, Academia Sinica, Taipei, Taiwan 11529, Republic of China}
\affiliation{Argonne National Laboratory, Argonne, Illinois 60439}
\affiliation{Institut de Fisica d'Altes Energies, Universitat Autonoma de Barcelona, E-08193, Bellaterra (Barcelona), Spain}
\affiliation{Istituto Nazionale di Fisica Nucleare, University of Bologna, I-40127 Bologna, Italy}
\affiliation{Brandeis University, Waltham, Massachusetts 02254}
\affiliation{University of California at Davis, Davis, California  95616}
\affiliation{University of California at Los Angeles, Los Angeles, California  90024}
\affiliation{University of California at San Diego, La Jolla, California  92093}
\affiliation{University of California at Santa Barbara, Santa Barbara, California 93106}
\affiliation{Instituto de Fisica de Cantabria, CSIC-University of Cantabria, 39005 Santander, Spain}
\affiliation{Carnegie Mellon University, Pittsburgh, PA  15213}
\affiliation{Enrico Fermi Institute, University of Chicago, Chicago, Illinois 60637}
\affiliation{Joint Institute for Nuclear Research, RU-141980 Dubna, Russia}
\affiliation{Duke University, Durham, North Carolina  27708}
\affiliation{Fermi National Accelerator Laboratory, Batavia, Illinois 60510}
\affiliation{University of Florida, Gainesville, Florida  32611}
\affiliation{Laboratori Nazionali di Frascati, Istituto Nazionale di Fisica Nucleare, I-00044 Frascati, Italy}
\affiliation{University of Geneva, CH-1211 Geneva 4, Switzerland}
\affiliation{Glasgow University, Glasgow G12 8QQ, United Kingdom}
\affiliation{Harvard University, Cambridge, Massachusetts 02138}
\affiliation{The Helsinki Group: Helsinki Institute of Physics; and Division of High Energy Physics, Department of Physical Sciences, University of Helsinki, FIN-00044, Helsinki, Finland}
\affiliation{Hiroshima University, Higashi-Hiroshima 724, Japan}
\affiliation{University of Illinois, Urbana, Illinois 61801}
\affiliation{The Johns Hopkins University, Baltimore, Maryland 21218}
\affiliation{Institut f\"{u}r Experimentelle Kernphysik, Universit\"{a}t Karlsruhe, 76128 Karlsruhe, Germany}
\affiliation{High Energy Accelerator Research Organization (KEK), Tsukuba, Ibaraki 305, Japan}
\affiliation{Center for High Energy Physics: Kyungpook National University, Taegu 702-701; Seoul National University, Seoul 151-742; and SungKyunKwan University, Suwon 440-746; Korea}
\affiliation{Ernest Orlando Lawrence Berkeley National Laboratory, Berkeley, California 94720}
\affiliation{University of Liverpool, Liverpool L69 7ZE, United Kingdom}
\affiliation{University College London, London WC1E 6BT, United Kingdom}
\affiliation{Massachusetts Institute of Technology, Cambridge, Massachusetts  02139}
\affiliation{Institute of Particle Physics, McGill University, Montr\'{e}al, Canada H3A~2T8; and University of Toronto, Toronto, Canada M5S~1A7}
\affiliation{University of Michigan, Ann Arbor, Michigan 48109}
\affiliation{Michigan State University, East Lansing, Michigan  48824}
\affiliation{Institution for Theoretical and Experimental Physics, ITEP, Moscow 117259, Russia}
\affiliation{University of New Mexico, Albuquerque, New Mexico 87131}
\affiliation{Northwestern University, Evanston, Illinois  60208}
\affiliation{The Ohio State University, Columbus, Ohio  43210}
\affiliation{Okayama University, Okayama 700-8530, Japan}
\affiliation{Osaka City University, Osaka 588, Japan}
\affiliation{University of Oxford, Oxford OX1 3RH, United Kingdom}
\affiliation{University of Padova, Istituto Nazionale di Fisica Nucleare, Sezione di Padova-Trento, I-35131 Padova, Italy}
\affiliation{University of Pennsylvania, Philadelphia, Pennsylvania 19104}
\affiliation{Istituto Nazionale di Fisica Nucleare, University and Scuola Normale Superiore of Pisa, I-56100 Pisa, Italy}
\affiliation{University of Pittsburgh, Pittsburgh, Pennsylvania 15260}
\affiliation{Purdue University, West Lafayette, Indiana 47907}
\affiliation{University of Rochester, Rochester, New York 14627}
\affiliation{The Rockefeller University, New York, New York 10021}
\affiliation{Istituto Nazionale di Fisica Nucleare, Sezione di Roma 1, University di Roma ``La Sapienza," I-00185 Roma, Italy}
\affiliation{Rutgers University, Piscataway, New Jersey 08855}
\affiliation{Texas A\&M University, College Station, Texas 77843}
\affiliation{Texas Tech University, Lubbock, Texas 79409}
\affiliation{Istituto Nazionale di Fisica Nucleare, University of Trieste/\ Udine, Italy}
\affiliation{University of Tsukuba, Tsukuba, Ibaraki 305, Japan}
\affiliation{Tufts University, Medford, Massachusetts 02155}
\affiliation{Waseda University, Tokyo 169, Japan}
\affiliation{Wayne State University, Detroit, Michigan  48201}
\affiliation{University of Wisconsin, Madison, Wisconsin 53706}
\affiliation{Yale University, New Haven, Connecticut 06520}


\author{D.~Acosta}
\affiliation{University of Florida, Gainesville, Florida  32611}

\author{T.~Affolder}
\affiliation{University of California at Santa Barbara, Santa Barbara, California 93106}

\author{T.~Akimoto}
\affiliation{University of Tsukuba, Tsukuba, Ibaraki 305, Japan}

\author{M.G.~Albrow}
\affiliation{Fermi National Accelerator Laboratory, Batavia, Illinois 60510}

\author{D.~Ambrose}
\affiliation{University of Pennsylvania, Philadelphia, Pennsylvania 19104}

\author{S.~Amerio}
\affiliation{University of Padova, Istituto Nazionale di Fisica Nucleare, Sezione di Padova-Trento, I-35131 Padova, Italy}

\author{D.~Amidei}
\affiliation{University of Michigan, Ann Arbor, Michigan 48109}

\author{A.~Anastassov}
\affiliation{Rutgers University, Piscataway, New Jersey 08855}

\author{K.~Anikeev}
\affiliation{Massachusetts Institute of Technology, Cambridge, Massachusetts  02139}

\author{A.~Annovi}
\affiliation{Istituto Nazionale di Fisica Nucleare, University and Scuola Normale Superiore of Pisa, I-56100 Pisa, Italy}

% ...


\collaboration{The CDF Collaboration}  \noaffiliation

\date{\today}
\begin{abstract}
We report the first measurements of inclusive $W$ and $Z$ cross sections
times leptonic branching ratios for $\ppbar$ collisions at $\sqrt{s} = 1.96$~TeV,
based on their decays to electrons and muons. The data correspond to an integrated 
luminosity of $72~\pbinv$ recorded with the CDF~detector at the Fermilab 
Tevatron.  We test $e$-$\mu$ universality in $W$ decays, and we measure the ratio 
of leptonic $W$ and $Z$ rates from which the leptonic branching fraction 
$B(W\goto\ell\nu)$ can be extracted as well as an indirect value for 
the total width of the $W$ and the CKM matrix element, $\Vcs$.
\end{abstract}
%
%
\pacs{13.38.Be,14.70.Fm,13.85.Qk,12.38.Qk,12.15.Ji}
\maketitle
%
\par
We report the first measurements of the inclusive production of $W$ and $Z$
bosons in $\ppbar$ collisions at the upgraded Run~II \fermilab\ \tevatron\ operated 
at $\sqrt{s} = 1.96$~TeV.  Measurements during Run~I at $\sqrt{s} = 1.8$~TeV have 
been reported by the \cdf\ and \dzero\ collaborations~\cite{runImeas}.
The data were collected with the \cdf\ detector and comprise $72.0\pm 4.3~\pbinv$.  
$W$ and $Z$ bosons are identified by their decays to electrons and muons,
from which we obtain the total rates $\sigma(\ppbar\goto W)\times B(W\goto\ell\nu)$
and $\sigma(\ppbar\goto Z)\times B(Z\goto\lplm)$~\cite{defineell}.
We test $e$-$\mu$ universality in $W$ decays using the ratio of inclusive $W$ 
production for the two lepton species.  We derive the leptonic 
branching ratio $B(W\goto\ell\nu)$ and an indirect value for the total 
$W$ width, $\Gammatot$, from the ratio of leptonic rates,
\begin{equation}
\label{def:R}
R \equiv  \frac{ \sigma(\ppbar\goto W)\times B(W\goto\ell\nu) }
                 { \sigma(\ppbar\goto Z)\times B(Z\goto\lplm) }  .
\end{equation}
Measurements of $\Gammatot$ test the Standard Model (SM), 
which predicts $\Gammatot$ in terms of electroweak parameters and 
the CKM matrix elements~\cite{renton}.
\par
The \cdf\ detector has been substantially upgraded since the end
of the previous data-taking period~\cite{tdr}. The central outer
tracker (COT) is a precision drift chamber which provides up to 
$96$~space-points for a track falling within its fiducial 
range $|\eta| < 1$~\cite{kinematics}.  The COT sense wires are 
arranged in eight `super-layers,' four of which provide axial 
coordinates and four of which provide stereo measurements. Precise 
track coordinates closer to the beam are provided by the silicon 
vertex detector.  The fiducial coverage of the central electromagnetic~(em)
and hadronic~(had) calorimeters has been extended to $|\eta| \sim 2.8$,
and the muon chambers provide coverage out to $|\eta| \sim 1$.
\par
The selection of candidate $W$ and $Z$ decays begins with the 
requirement of a high-$p_T$ lepton.
Electrons are identified on the basis of their electromagnetic showers.
We require an energy cluster in a well-instrumented region of
the calorimeter with $\ET > 25$~GeV, matched to a single track 
with $\pt > 10$~GeV that extrapolates to the position of the cluster
at a depth corresponding to shower maximum. The ratio $\ET/\pt$ must be 
less than~$2$, and the energy in the hadronic calorimeter must be relatively 
small: $\Ehad/\Eem < 0.055 + 0.00045\times \Eem$.  The shape of the 
shower must be consistent with that observed from test-beam data.  
Beyond the central tracker coverage, $|\eta| > 1$, only calorimetry
is used to identify electrons.
\par
Muons are identified on the basis of a track segment (`stub') 
reconstructed in the muon chambers with positions well-matched
to the extrapolation of a single track.  We require $\pt > 20$~GeV, 
and energy depositions in the calorimeters consistent with those
expected from a minimum-ionizing particle.
\par
Requirements on the reconstructed track are common to both the
electron and muon selections.  At least three axial and three stereo 
COT super-layers must have seven hits or more.  Not all the lepton
tracks have hits from the silicon vertex detector, so for the sake
of uniformity in the $p_T$ measurement, we drop these hits and constrain
the track fit to the transverse beam profile.  For muons, we apply a cut 
on the $\chi^2$ for the track fit to eliminate kaons and pions which 
have decayed in flight. The coordinate of the lepton along the beam line
must fall within $60$~cm of the center of the detector to ensure a good 
energy measurement in the calorimeter.  This requirement eliminates 
$\sim 5\%$ of the data, according to studies with minimum-bias events.
\par
After the selection of high-$p_T$ leptons, we establish the $W$ and $Z$ samples.
For $W\goto\ell\nu$ candidates, we require $\MET > 25$~GeV ($20$~GeV) 
in the electron (muon) channel as evidence for the neutrino.
In the muon channel, events with any second charged particle ($p_T > 10$~GeV) are 
rejected as potential background from $Z\goto\mpmm$.  For $Z$ candidates, a second 
electron is required in the electron channel, and a second 
charged particle in the muon channel.  These must pass the same $E_T$ and
$p_T$ cuts as the first lepton. The identification requirements are looser 
for the second lepton in order to maintain good efficiency for these events.
For example, for loose electrons in the central region, an associated track 
is required, but the requirements on $E_T/p_T$, of the match of the track
to the center of the cluster, and on the lateral shower shape are dropped.
\par
For $W\rightarrow e\nu$ candidates we accept electrons reconstructed
in the central calorimeter, which corresponds to $|\eta| \lesssim 1$.  
For $Z\rightarrow \epem$ candidates, one electron must come
from the central calorimeter, while the second can come from 
the forward region, which extends out to $|\eta| = 2.8$.
\par
A significant background comes from leptons from the decays of heavy-flavor 
hadrons, which can be reduced by requiring that the lepton be isolated.
We require that the calorimetric energy not associated with the lepton in a cone 
of $\Delta R = 0.4$ around the lepton must be no more than 10\% of the energy 
of the lepton~\cite{deltaR}.
\par
Cosmic-ray muons contaminate the muon samples.  We use the timing 
capabilities of the COT to reject events with two muon tracks, one 
of which travels from outside of the COT toward the beam pipe.
We also require that the muon tracks pass close to the beam line,
within distances less than $0.02$~cm ($0.2$~cm) 
for tracks with (without) silicon hits.
\par
The kinematic and geometric acceptance is estimated using the \pythia~6.203 event 
generator~\cite{pythia} and a full simulation of the CDF detector based on 
the \geant\ simulation package~\cite{geant}.  The key quantity is the boson 
rapidity, $y$, so we extract the acceptance $A(y)$ from the simulation and 
convolve it with a NNLO calculation of $d\sigma/dy$~\cite{NNLO}, which depends 
on the parton distribution functions (PDF's).  We compute the central values
of the acceptance using the \mrstpdf\ PDF's~\cite{mrst}. We estimate the 
uncertainties using the eigenvector basis sets for \cteqsix~\cite{cteq6}, 
and obtain 1.3\%~for $W\rightarrow e\nu$ and~$\mu\nu$, and 0.7\%~for
$Z\rightarrow\epem$ and 2.1\%~for $Z\rightarrow\mpmm$.These are roughly a factor 
two larger than what we obtained using the \mrstpdf\ error PDF's.
\par
The amount of material an electron passes through is known to 
$10\%$--$30\%$, depending on $\eta$: this
contributes $\lesssim 1\%$ to the acceptance uncertainty for the
electron channels.  The energies of hadronic jets recoiling against the
$W$ bosons enter the calculation of $\MET$.  We test the accuracy of the
simulation using a $\chi^2$~test with scale factors and offsets for the
components of these energies which are parallel and perpendicular to the
lepton momentum vector.  Taking a variation corresponding to
$\Delta\chi^2 = 9$, we infer systematic uncertainties of about $0.3\%$.
The energy and momentum scales for the leptons are treated in a similar
manner, resulting in an uncertainty of $0.2\%$--$0.3\%$.   Finally, we vary
the parameters of the \pythia\ model which influence the boson $p_T$
distribution, as allowed by $\chi^2$~tests with the Run~I 
measurement~\cite{runImeas}, and find the uncertainty on the acceptance 
is negligible.
\par
Lepton reconstruction, identification, isolation and trigger efficiencies
are measured directly with the data.
We use $Z\goto\lplm$ decays in which the standard cuts are applied to 
one lepton and the second candidate lepton is tested to see 
whether it passes the given criteria.  The statistical uncertainties are 
below 1\% and studies with the simulation show negligible bias with 
respect to $W$ decays. The cuts to eliminate cosmic rays remove a very 
small fraction of signal events in the muon channel, as measured 
by applying the cuts to $W\goto e\nu$ and $Z\goto\epem$ events.
The track-reconstruction efficiency is measured using a trigger demanding
an energetic calorimeter cluster which provides a sample of $W\rightarrow e\nu$
events independent of the tracking.  The efficiency for rejecting
$Z\rightarrow \mpmm$ events in the $W\rightarrow\mu\nu$ channels is estimated
using the simulation.
\par
Backgrounds fall into three categories: multi-jet events with no $W$ or $Z$
bosons, weak-boson backgrounds ($Z\goto\lplm$ and $W\goto\tau\nu$ appearing 
in $W\goto\ell\nu$, and $Z\goto\tptm$ and $W\goto\ell\nu$ appearing in 
$Z\goto\lplm$) and non-collision  background, primarily cosmic rays.   
Estimates of these backgrounds are summarized in Table~\ref{backgrounds}.
\par
The multi-jet background is estimated with the data.  Such events are
characterized by a significant energy in the cone around the lepton
and a small $\MET$, with tails in both quantities.  We assume that
these tails are not correlated, and estimate the number of background 
events by comparing to control regions with either low $\MET$ or high 
energy in the isolation cone, after taking into account the $W$ events
which fall in the control regions.  We vary the cuts on 
$\MET$ and isolation which define the control regions, and 
then estimate changes in a manner reproduced by the simulation.
We assign a systematic uncertainty of 50\% for this background estimate.
\par
The weak-boson backgrounds are obtained using the simulation to compute
the acceptance relative to that of the signal.  To normalize the
contribution of $Z\goto\lplm$ backgrounds to $W\goto\ell\nu$ channels,
we use the theoretical value for the ratio of cross sections, with
an uncertainty corresponding to previous measurements of $W$ and $Z$ 
cross sections~\cite{runImeas}.  Backgrounds from di-boson and $t\bar{t}$ 
production are negligible.
\par
For estimating the cosmic-ray contamination in the $W\goto\mu\nu$ sample, 
we use the azimuthal distribution of muon chamber hits opposite the 
high-$p_T$ muon. For the $Z\goto\mpmm$ sample, we use the distribution 
of the cosine of the angle between the two muon tracks. In both cases, 
the contribution from cosmic rays is very small.

%
\begin{table}
\caption[.]{\label{backgrounds}
Background estimates.}
\begin{ruledtabular}
\begin{tabular}{lcccc}
 & \multicolumn{4}{c}{channel} \cr
category & $W\rightarrow e\nu$ & $W\rightarrow\mu\nu$ & 
           $Z\rightarrow e^+e^-$ & $Z\rightarrow\mu^+\mu^-$ \cr
\hline
multi-jet                  & $587\pm299$ & $220\pm 111$  & 
                             $41\pm 18$ &  $0^{+1}_{-0}$ \cr
$Z\rightarrow\ell^+\ell^-$ & $317\pm14$  & $1739\pm 75$ &  -  &  -  \cr
$Z\rightarrow\tau^+\tau^-$ & negl. & negl. & $3.7\pm 0.4$ & $1.5\pm 0.3$ \cr
$W\rightarrow\tau\nu$      & $752\pm17$  & $998\pm 31$ & negl. & negl. \cr
$W\rightarrow\ell\nu$      &  -  &  -  & $16.8 \pm 2.8$ & negl. \cr
cosmic rays                & negl. & $33\pm 23$ & negl. & $12\pm 12$ \cr
\end{tabular}
\end{ruledtabular}
\end{table}
%

The uncertainty on the luminosity measurement is 6.0\%, of which 4.4\% comes
from the acceptance and operation of the luminosity monitor and 4.0\% comes 
from the calculation of the total $\ppbar$ cross section~\cite{lumdoc}.  
%
\par
\begin{figure}
\begin{center}
\includegraphics[width=0.4\textwidth]{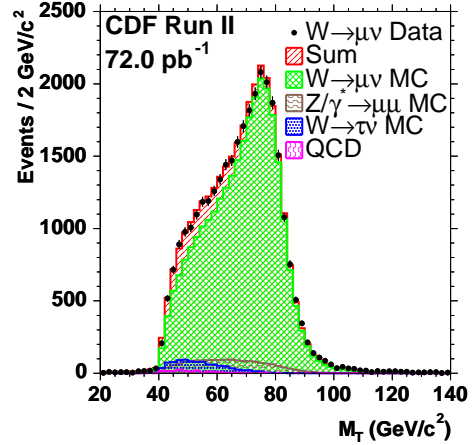}\\
\includegraphics[width=0.4\textwidth]{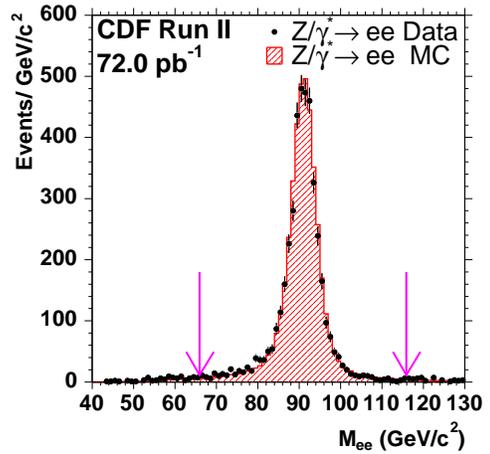}
\caption[.]{\label{MTplot}
Distributions of $M_T$ for $W\rightarrow\mu\nu$ (top) and
$M_{\epem}$ for $Z\gamma^* \rightarrow e^+e^-$ (bottom).
The arrows in the $M_{\epem}$ distribution define the mass window.
}
\end{center}
\end{figure}

We compute the transverse mass of each candidate $W$~decay:
$M_T \equiv \sqrt{E_T\MET - (E_x\METx + E_y\METy)}$, where $E_x$ and $E_y$
are measured with the calorimeter for electrons, and with the COT for muons.
In the $Z\goto\lplm$ channels, we compute the invariant mass of the lepton pair,
and count the candidates in the mass window
$66~{\mathrm{GeV}} < M_{\ell^+\ell^-} < 116~{\mathrm{GeV}}$.
The cross sections $\sigma(\ppbar\rightarrow Z/\gamma^* \rightarrow\lplm)$ 
reported here include the contributions from virtual photon exchange.  
Distributions for $W\rightarrow\mu\nu$ and $Z\rightarrow e^+e^-$ are 
shown in Fig.~\ref{MTplot}, which demonstrate that the simulations
match the data well.
%
\par
The measurement of the cross section requires the number of events
selected for the given luminosity, and the estimates of the acceptance, 
efficiencies, and backgrounds.  A summary of these quantities and the
inferred cross sections is given in Table~\ref{inputs}.
\par
We test $e$-$\mu$ universality in $W$ decays by taking ratios of the
$W$~cross sections.  Many uncertainties cancel in this ratio, and 
we find for the ratio of $W$-$\ell$-$\nu$ couplings:
$ g_\mu/g_e = 0.998 \pm 0.004_\stat \pm 0.011_\syst $.
\par
Since there is no sign of non-universality, we combine our
measurements taking correlated uncertainties into account, and obtain
\begin{eqnarray*}
 &\sigma\times B&(\ppbar\goto W\goto\ell\nu) = \cr
    & &  2775 \pm 10_\stat \pm 53_\syst \pm 167_\lumerr {\mathrm{~pb}} \cr
 &\sigma\times B&(\ppbar\goto Z/ \gamma^* \goto\lplm) = \cr
    & & 254.9 \pm 3.3_\stat \pm 4.6_\syst \pm 15.2_\lumerr  {\mathrm{~pb}} .
\end{eqnarray*}
Agreement with SM predictions~\cite{theory,stirling} is good.
\par
We compute the ratio, $R$ (Eq.~\ref{def:R}), taking all correlations 
among channels into account, and obtain 
$  R = 10.92 \pm 0.15_\stat \pm 0.14_\syst $,
after correcting for virtual photon exchange
(we multiply the measured $Z/\gamma^*$ cross section by a factor
$1.004\pm 0.001$.) Using the measured value 
$B(Z\goto\lplm) = 0.033658\pm 0.000023$~\cite{rpp} and a theoretical 
calculation of the ratio of production cross sections~\cite{stirling}, 
we extract the leptonic branching ratio
$ B(W\goto\ell\nu) = 0.1089 \pm 0.0022 $.
\par
Using the theoretical value for the leptonic partial width, 
$\Gamma(W\goto\ell\nu) = 226.4\pm 0.3$~MeV~\cite{rpp}, we extract 
the total width of the $W$ boson:
$  \Gammatot = 2079 \pm 41 {\mathrm{~MeV}} $
which can be compared to the SM value, $2092.1 \pm 2.5~{\mathrm{MeV}}$
and the current world average, $2118 \pm 42~{\mathrm{MeV}}$~\cite{rpp}.
Alternatively, rather than using the measured value for $B(Z\goto\lplm)$,
we can use the SM value for $\Gamma(Z\goto\lplm)$ and extract the ratio of
total widths: $\Gammatot / \Gamma_Z^{\mathrm{tot}} = 0.833 \pm 0.017$.
\par
Finally, in the SM, the total width $\Gammatot$ depends on electroweak 
parameters and certain CKM matrix elements, which we can constrain
using $\Gammatot$~\cite{renton}. Using world average values~\cite{rpp} 
for all CKM matrix elements except $\Vcs$, we derive $\Vcs = 0.967\pm 0.030$.
\par
\begin{table*}
\caption[.]{\label{inputs}
Number of selected events, and estimated acceptance, efficiency,
and expected number of background events.  Cross sections are reported in pb, with
a statistical and systematic uncertainty. A common uncertainty due 
to the luminosity measurement is $166$~pb ($15$~pb) for the $W$ ($Z$) channels.}
\begin{ruledtabular}
\begin{tabular}{lcccc}
 & \multicolumn{4}{c}{channel} \cr
category & $W\rightarrow e\nu$ & $W\rightarrow\mu\nu$ & 
           $Z/\gamma^*\rightarrow e^+e^-$ & $Z/\gamma^*\rightarrow\mu^+\mu^-$ \cr
\hline
$N$ candidates        & 37584 & 31722 & 4242 & 1785 \cr
acceptance            & $0.2397\pm0.0039$ & 
                        $0.1970\pm0.0027$ & 
                        $0.3182\pm0.0041$ & 
                        $0.1392\pm0.0030$ \cr
efficiency            & $0.749\pm0.009$ & 
                        $0.732\pm0.013$ & 
                        $0.713\pm0.012$ & 
                        $0.713\pm0.015$ \cr
background            & $1656\pm 300$ & 
                        $2990\pm 140$ & 
                        $62 \pm 18$ & 
                        $13 \pm 13$ \cr
\hline
cross section (pb)    & $2780\pm 14\pm 60$ & 
                        $2768\pm 16\pm 64$ & 
                        $255.8\pm 3.9\pm 5.5$ & 
                        $248.0\pm 5.9\pm 7.6$ \cr
\end{tabular}
\end{ruledtabular}
\end{table*}
%
\par
\begin{acknowledgments}
We appreciate the help we received from William~Stirling and Lance~Dixon.
We thank the Fermilab staff and the technical staffs of the participating 
institutions for their vital contributions. This work was supported by the 
U.S. Department of Energy and National Science Foundation; the Italian 
Istituto Nazionale di Fisica Nucleare; the Ministry of Education, Culture, 
Sports, Science and Technology of Japan; the Natural Sciences and Engineering 
Research Council of Canada; the National Science Council of the Republic of 
China; the Swiss National Science Foundation; the A.P. Sloan Foundation; 
the Research Corporation; 
the Bundesministerium f\"ur Bildung und Forschung, Germany; the Korean 
Science and Engineering Foundation and the Korean Research Foundation; 
the Particle Physics and Astronomy Research Council and the Royal Society, UK; 
the Russian Foundation for Basic Research; the Comision Interministerial 
de Ciencia y Tecnolog\`ia, Spain; in part by the European Community's 
Human Potential Programme under contract HPRN-CT-20002, Probe for New 
Physics; and by the Research Fund of Istanbul University Project 
No.~1755/21122001. 
\end{acknowledgments}

%


\begin{thebibliography}{99}
\bibitem{kinematics}
Let $\theta$ be the polar angle with respect to the proton beam axis,
and $\phi$ the azimuthal angle.
The {\sl pseudo-rapidity} is $\eta \equiv -\ln(\tan(\theta/2))$.
The {\sl transverse momentum}, $p_T$, is the component of the momentum
projected onto the plane perpendicular to the beam axis.
The {\sl transverse energy}, $E_T$, of a shower or calorimeter tower
is $E\sin\theta$.  The {\sl missing energy}, $\MET$, is the vector which points 
opposite the vector sum of all calorimeter energy: ${\MET}_x = -\sum_i E_{T,i}\cos\phi_i$, 
where the sum extends over all towers, $E_{T,i}$ is the transverse energy of 
the $i^{\mathrm{th}}$ tower. A similar expression holds for the $y$-component.
\end{thebibliography}
\end{document}